\documentclass[a4paper,aps,prl,reprint,preprintnumbers,amsmath,amssymb,superscriptaddress]{revtex4-1}

\usepackage{amsmath}
\usepackage{verbatim}
\usepackage{graphicx}
\usepackage{color}
\usepackage{ragged2e}
\usepackage{wrapfig}
\usepackage{multibib}
\usepackage[left]{lineno}

\nolinenumbers

 \newcommand{\ket}[1]{\left|#1\right\rangle}

\newcommand{\iqoqi}{\affiliation{Institute for Quantum Optics and Quantum Information of the
Austrian Academy of Sciences, 6020 Innsbruck, Austria}}
\newcommand{\exphys}{\affiliation{Institute for Experimental Physics, University of Innsbruck, 6020 Innsbruck, Austria}}
\newcommand{\thphys}{\affiliation{Institute for Theoretical Physics, University of Innsbruck, 6020 Innsbruck, Austria}}
\newcommand{\garching}{\affiliation{Physik Department, Technische Universit\" ät M\" unchen, 85747 Garching, Germany\\ *These authors contributed equally to this work.}}

\definecolor{mygreen}{rgb}{0,0.5,0}
\definecolor{myblue}{rgb}{0,0,0.75}
\definecolor{mymagenta}{cmyk}{0,1,0,0.12}

\begin{document}

\title{Real-time dynamics of lattice gauge theories with a few-qubit quantum computer}

\author{Esteban A. Martinez} \thanks{These authors contributed equally to this work.} \exphys
\author{Christine Muschik} \thanks{These authors contributed equally to this work.} \iqoqi \thphys 
\author{Philipp Schindler} \exphys
\author{Daniel Nigg} \exphys
\author{Alexander Erhard} \exphys
\author{Markus Heyl}\iqoqi \garching
\author{Philipp Hauke} \iqoqi \thphys 
\author{Marcello Dalmonte} \iqoqi \thphys 
\author{Thomas Monz} \exphys
\author{Peter Zoller} \iqoqi \thphys 
\author{Rainer Blatt} \exphys \iqoqi

\maketitle

\textbf{Gauge theories are fundamental to our understanding of interactions between the elementary constituents of matter as mediated by gauge bosons~\cite{Gattringer2010,calzetta_book}. However, computing the real-time dynamics in gauge theories is a notorious challenge for classical computational methods.
In the spirit of Feynman's vision of a quantum simulator~\cite{Feynman82,Nori2014}, this has recently stimulated theoretical effort to devise schemes for simulating such theories on engineered quantum-mechanical devices, with the difficulty that gauge invariance and the associated local conservation laws (Gauss laws) need to be implemented~\cite{Preskill2012,Wiese:2013kk,Zohar2015}.
Here we report the first experimental demonstration of a digital quantum simulation of a lattice gauge theory, by realising  1+1-dimensional quantum electrodynamics (Schwinger model~\cite{Schwinger1,KogutSusskindFormulation}) on a few-qubit trapped-ion quantum computer.
We are interested in the real-time evolution of the Schwinger mechanism~\cite{Hebenstreit2013,Kasper2015}, describing the instability of the bare vacuum due to quantum fluctuations, which manifests itself in the spontaneous creation of electron-positron pairs. To make efficient use of our quantum resources, we map the original problem to a spin model by eliminating the gauge fields~\cite{Encoding} in favour of exotic long-range interactions, which have a direct and efficient implementation on an ion trap architecture~\cite{Blatt2012}. We explore the Schwinger mechanism of particle-antiparticle generation by monitoring the mass production and the vacuum persistence amplitude. Moreover, we track the real-time evolution of entanglement in the system, which illustrates how particle creation and entanglement generation are directly related. Our work represents a first step towards quantum simulating high-energy theories with  atomic physics experiments, the long-term vision being the extension to real-time quantum simulations of non-Abelian lattice gauge theories.}

 \begin{figure}
\centering
\includegraphics[width=\columnwidth]{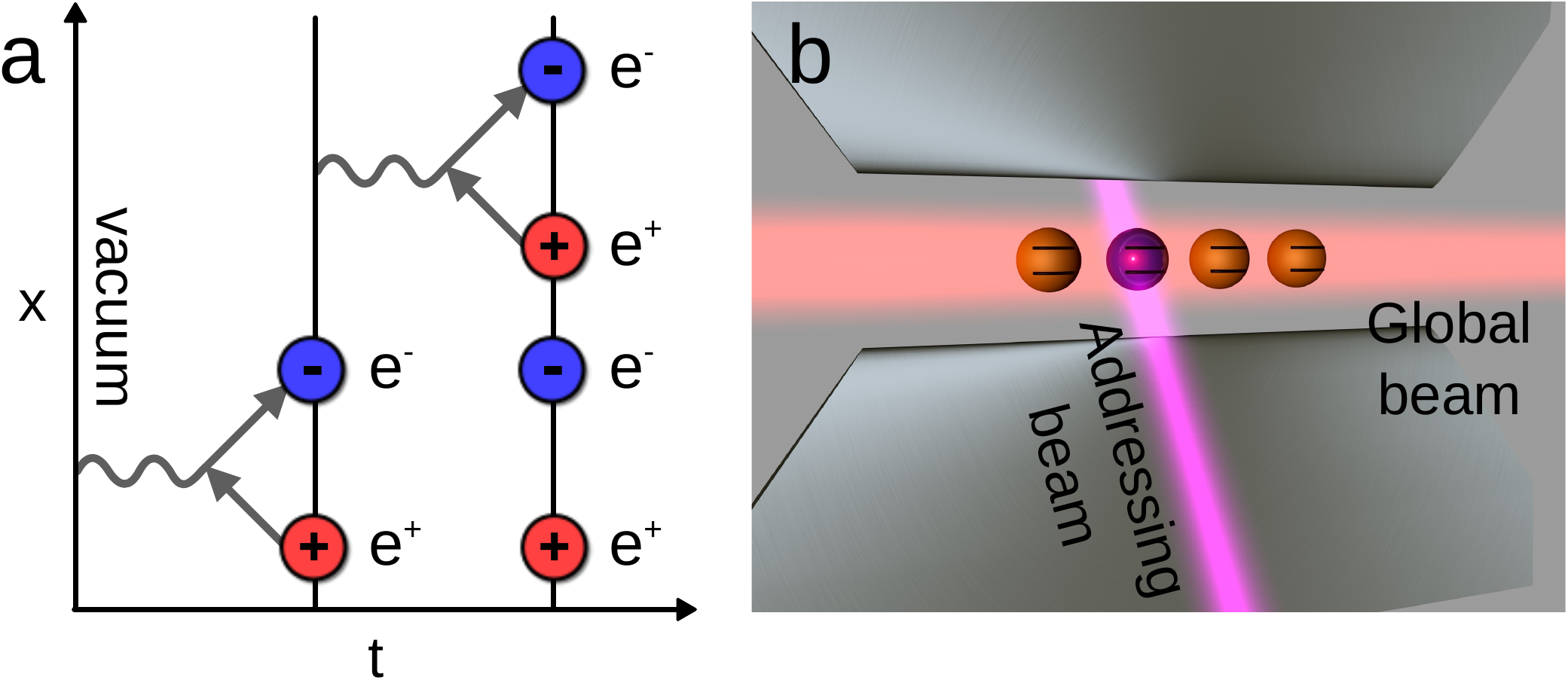}
\caption{(a) The instability of the vacuum due to quantum fluctuations is one of the most fundamental effects in gauge theories. We simulate the coherent real time dynamics of particle-antiparticle creation by realising the Schwinger model (one-dimensional quantum electrodynamics) on a lattice, as described in the main text. (b) The experimental setup for the simulation consists of a linear Paul trap, where a string of $^{40}$Ca$^+$ ions is confined. The electronic states of each ion encode a spin $|\!\!\uparrow\rangle$ or $|\!\!\downarrow\rangle$; these can be manipulated using laser beams (see Methods for details).\label{Fig_1} }
\end{figure} 
 
Small-scale quantum computers exist today in the laboratory as programmable quantum devices~\cite{ReviewQuantumComputers}. In particular, trapped-ion quantum computers~\cite{Blatt2012} provide a platform allowing a few hundred coherent quantum gates on a few qubits, with a clear roadmap towards scaling up these devices~\cite{Nori2014,Monroe2014}.
This provides the tools for universal digital quantum simulation~\cite{Lanyon2011}, where the time evolution of a quantum system is approximated as a stroboscopic sequence of quantum gates~\cite{Lloyd96}. 
Here we show how this quantum technology can be used to simulate the real time dynamics of a minimal model of a lattice gauge theory, realising the Schwinger model~\cite{Schwinger1,KogutSusskindFormulation} as a 1D quantum field theory with a chain of trapped ions (see Fig.~\ref{Fig_1}). 

 \begin{figure*}
  \centering
  \includegraphics[width=\linewidth]{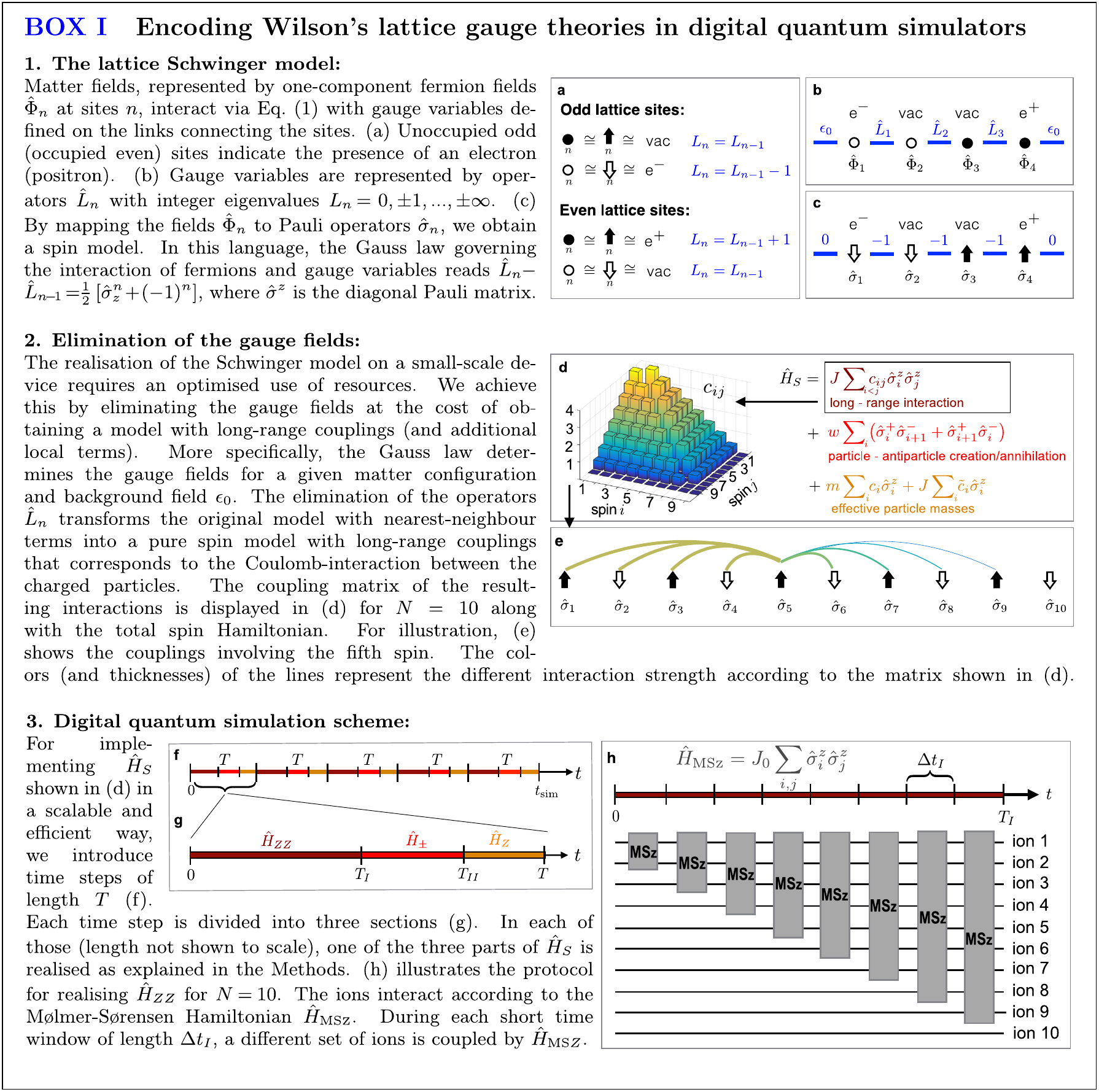}
\end{figure*}

Our few-qubit demonstration is a first step towards simulating real time dynamics in gauge theories, which is fundamental for the understanding of many physical phenomena including the thermalisation after heavy-ion collisions and pair creation studied at high-intensity laser facilities~\cite{Narozhny2014,Wiese:2013kk}. While existing classical numerical methods such as Quantum Monte Carlo have been remarkably successful for describing equilibrium phenomena, no systematic techniques exist to tackle the dynamical long-time behaviour of all but very small systems. In contrast, quantum simulations aim at the long-term goal to solve the specific yet fundamental class of problems that currently cannot be tackled by these classical techniques. The digital approach we employ here is based on the Hamiltonian formulation of gauge theories~\cite{KogutSusskindFormulation}, and enables direct access to the system wave-function. As we show below, this allows us to investigate entanglement generation during particle-antiparticle production, emphasising a novel perspective on the dynamics of the Schwinger mechanism~\cite{calzetta_book}. 

Digital quantum simulations described in the present work are conceptually different from, and fundamentally more challenging than, previously reported condensed matter-motivated simulations of spin and Hubbard-type models~\cite{Nori2014,Barends2015,Wallraff2015}. In gauge theories, local symmetries lead to the introduction of dynamical gauge fields obeying a Gauss law~\cite{Wiese:2013kk}. 
Formally, this crucial feature is described by local symmetry generators $\{\hat{G}_i\}$ that commute with the Hamiltonian of the system $[\hat{H},\hat{G}_i]=0$ and restrict the dynamics to a subspace of physical states $\hat{G}_i|\Psi_{\text{\tiny{physical}}}\rangle=0$. 
Realising such a constrained dynamics  on a quantum simulator is demanding and has been the focus of theoretical research~\cite{Wiese:2013kk,Zohar2015,Zohar2012,Tagliacozzo:2012kq,Banerjee2012,Hauke2013b,Kasper2015}. Instead, to optimally use the finite resources represented by a few qubits of existing quantum hardware, we encode the gauge degrees of freedom in the form of a long-range interaction between the fermions, which can be implemented very efficiently on our experimental platform. This allows us to explore quantum simulation of coherent real-time dynamics with four qubits, exemplified here by the creation of electron-positron pairs (see Fig.~\ref{Fig_1}). 

To this end, we experimentally study the Schwinger model, which describes quantum electrodynamics in one dimension. This model is extensively used as a testbed for lattice gauge theories as it shares many important features with quantum chromodynamics, including confinement, chiral symmetry breaking, and a topological theta vacuum \cite{Wiese:2013kk}. In the Kogut--Susskind Hamiltonian formulation of the Schwinger model~\cite{Schwinger1,KogutSusskindFormulation},

\begin{align}\label{Eq_LatticeHamiltonian}
\hat{H}_{\text{lat}}&=-i w\sum_{n=1}^{N-1}\left[\hat{\Phi}^{\dag}_ne^{i\hat{\theta}_n}\hat{\Phi}_{n+1}-\text{H.c.}\right]\\
& \quad +J\sum_{n=1}^{N-1} \hat{L}_n^2+m \sum_{n=1}^{N}(-1)^n \hat{\Phi}^{\dag}_n\hat{\Phi}_n \nonumber
\end{align}
describes the interaction of fermionic field operators $\hat{\Phi}_n$ at sites $n=1\dots N$ with gauge fields that are represented by the canonically commuting operators $[\hat{\theta}_n,\hat{L}_m]=i\delta_{n,m}$.  $\hat{L}_n$ and $\hat{\theta}_n$ correspond to the electromagnetic field and vector potential on the connection between sites $n$ and $n+1$. The latter can be eliminated by a gauge transformation (see Methods). The fields $\hat{\Phi}_n$ represent Kogut--Susskind fermions (see Box~I(a)), where the presence of an electron (positron) is mapped to an occupied odd (unoccupied even) lattice site, allowing for a convenient incorporation of particles and antiparticles in a single fermion field. Accordingly, the third term in Eq.~(\ref{Eq_LatticeHamiltonian}), representing the rest mass $m$, obtains a staggered sign. The first term corresponds to the creation and annihilation of particle-antiparticle pairs, and the second term reflects the energy stored in the electric field. Their energy scales $w=\frac{1}{2a}$ and $J=\frac{g^2 a}{2}$ depend on the lattice spacing $a$ and the fermion-light coupling constant $g$. Throughout the paper, we use natural units $\hbar=c=1$. Therefore, $a$ and $t$ have the dimension of length, while $w$, $J$, $m$ and $g$ have the dimension of inverse length.

For realising the model using trapped ions, we map the fermionic operators $\hat{\Phi}_n$ to spin operators (see Box~I(a)) by a Jordan--Wigner transformation~\cite{Encoding}, which converts the short-range hopping interactions in Eq.~(\ref{Eq_LatticeHamiltonian}) into nearest-neighbour spin flip terms. In this formulation, the Gauss law takes the form $\hat{L}_n-\hat{L}_{n-1}=\frac{1}{2}\left[\hat{\sigma}_n^z+(-1)^n\right]$, which is the lattice version of the continuum law $\nabla E=\rho$, where $\rho$ is the charge density.  As illustrated in Box~I(c), the Gauss law completely determines the electric fields for a given spin configuration and choice of background field. Following Ref.~\cite{Encoding}, we use this constraint to eliminate the operators $\hat{L}_n$ from the dynamics, adapting a scheme that has previously proven advantageous for numerical calculations~\cite{Banuls} to a quantum simulation experiment, where gauge invariance is fulfilled by construction.

\begin{figure}[h]
\centering
\includegraphics[width=0.95\columnwidth]{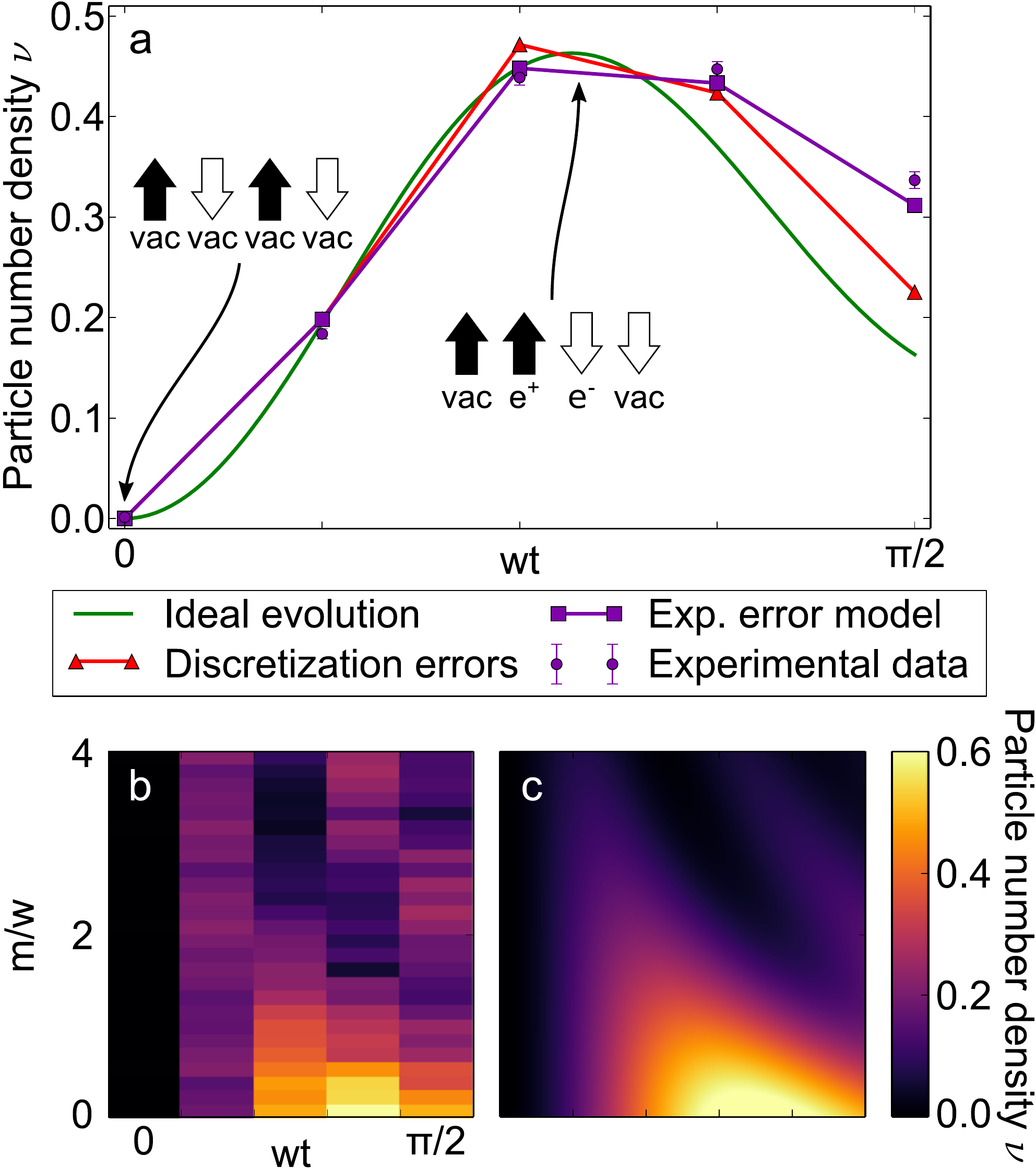}
\caption{(a) Time evolution of the particle number density $\nu$, starting from the bare vacuum. We show the ideal evolution under the Schwinger Hamiltonian $\hat{H}_S$ shown in Box I(d), the ideal evolution using discrete time steps (see Box~I), the expected evolution including an experimental error model (see Methods) and the actual postselected experimental data (see Methods) for electric field energy $J = w$ and particle mass $m = 0.5 \, w$ (see Eq.~(\ref{Eq_LatticeHamiltonian})). The insets show the initial state of the simulation, corresponding to the bare vacuum with particle number density $\nu \! = \! 0$, as well as one example of a state containing one pair, i.e.\ a state with $\nu = 0.5$. (b) Experimental data and (c) theoretical prediction for the evolution of the particle number density $\nu$ as a function of the dimensionless time $w t$ and the dimensionless particle mass $m/w$, with $J = w$.\label{Fig_2}}
\end{figure}

\begin{figure}[h]
\centering
\includegraphics[width=0.95\columnwidth]{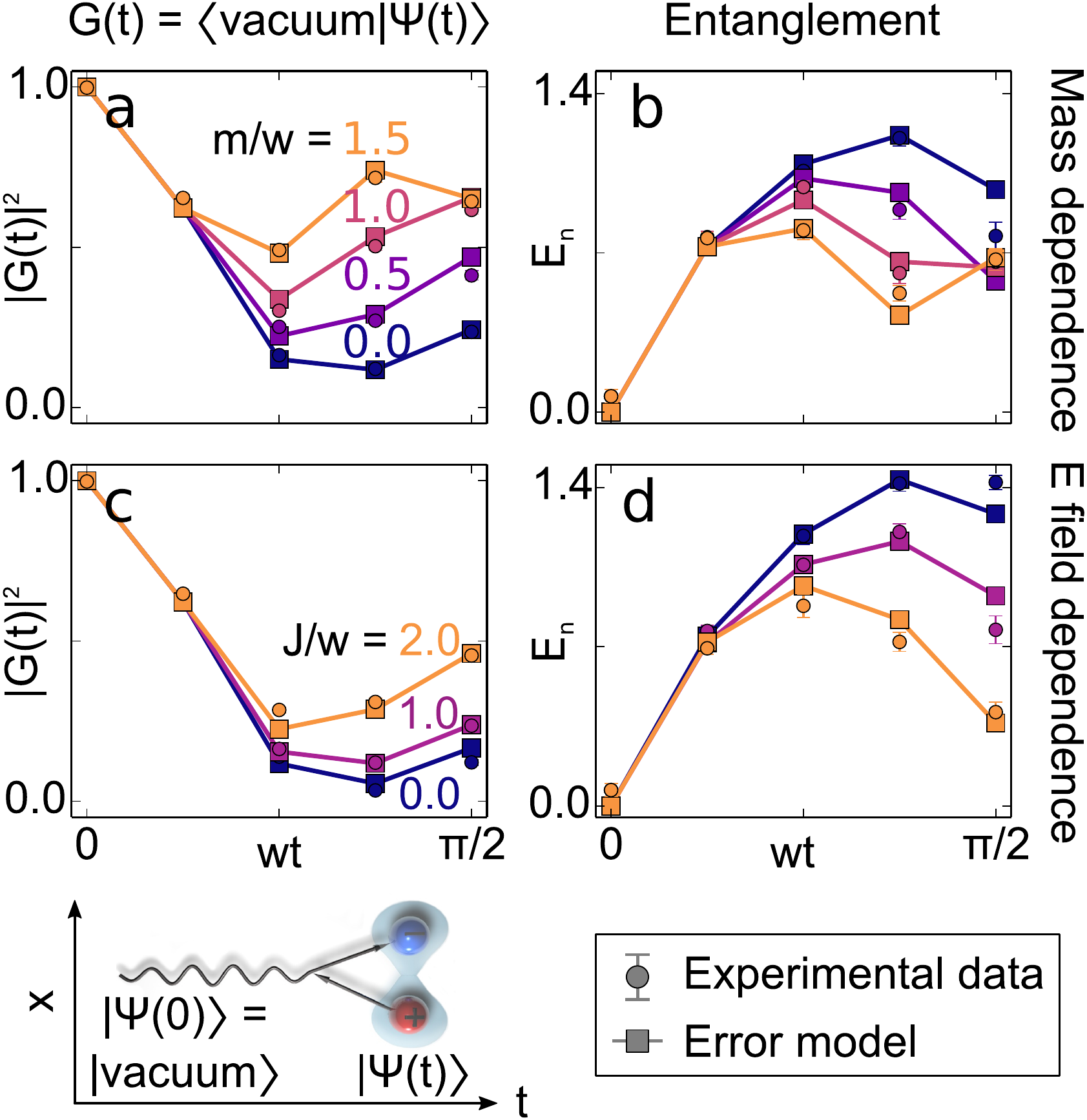}
\caption{Time evolution of the square of the vacuum persistence amplitude $|G(t)|^2$ (the Loschmidt echo), which quantifies the decay of the unstable vacuum and  the logarithmic negativity $E_n$ as a measure of the entanglement between the left and right half of the system. Panels (a) and (b) show the time evolution of these quantities for different values of the particle mass $m$ and fixed electric field energy $J=w$, where $w$ is the rate of particle-antiparticle creation and annihilation (compare Eq.~(\ref{Eq_LatticeHamiltonian})). Panels (c) and (d) show how the time evolution of $|G(t)|^2$ and $E_n$ changes for different values of $J$ and fixed particle mass $m=0$. Circles correspond to the experimental data and squares connected by solid lines to the expected evolution assuming an experimental error model explained in the Methods.\label{Fig_3}}
\end{figure}

The elimination of the gauge fields maps the original problem to a spin model with long-range interactions that reflect the Coulomb interactions between the simulated particles. This allows for a very efficient use of resources, since $N$ spins can be used to simulate $N$ particles and their accompanying $N\!-\!1$ gauge fields. However, as shown in Box~I(d), the required couplings and local terms have a very unusual distance and position dependence. The challenge has thus been moved from engineering a constrained dynamics of $2N\!-\!1$ quantum systems on a gauge-invariant Hilbert space to the realisation of an exotic and asymmetric interaction of $N$ spins.

Our platform is ideally suited for this task, since long-range interactions and precise single qubit operations are available in trapped-ion systems. These capabilities allow us to realise the required interactions by means of a digital quantum simulation scheme ~\cite{Lloyd96}. To this end, the desired Hamiltonian, $H=\sum_{k=1}^K H_k$, is split into $K$ parts that can be directly implemented and are applied separately in subsequent time windows. By repeating the sequence multiple times, the resulting time evolution of the system $U(t)$ closely resembles an evolution where the individual parts of the Hamiltonian act simultaneously, as can be shown using the Suzuki-Lie-Trotter expansion,
\begin{eqnarray*}
U(t)=e^{-i \hat{H} t}=\text{lim}_{n\rightarrow\infty}\left(\bigotimes_{k=1}^K e^{-i\hat{H}_kt/n}\right)^n.
\end{eqnarray*}
Our scheme is depicted in Box~I(f,g,h). It allows for a very efficient realisation of the required dynamics and implements the coupling matrix shown in Box~I(d,e) with a minimal number of time steps scaling only linearly in the number of sites $N$. The scheme is therefore scalable to larger systems. A discussion of finite size effects can be found in the Methods.

We realise the simulation in a quantum information processor based on
a string of $^{40}$Ca$^+$ ions confined in a macroscopic linear
Paul trap (see Fig. \ref{Fig_1}(b)). There, each qubit is encoded in the electronic states
$\mid\downarrow\rangle = 4 S_{1/2}$ (with magnetic quantum number $m = -1/2$),
$\mid\uparrow\rangle= 3 D_{5/2}$ ($m = -1/2$) of a single ion. The
energy difference between these states is in the optical domain, so the state of the qubit can be manipulated using laser light
pulses. More specifically, a universal set of high-fidelity quantum
operations is available, consisting of collective rotations around the
equator of the Bloch sphere, addressed rotations around the Z axis and
entangling M{\o}lmer-S{\o}rensen (MS) gates~\cite{MSgates}. With a sequence of
these gates, arbitrary unitary operations can be implemented~\cite{Schindler2013}. Thus, we are able to simulate any Hamiltonian evolution, and in particular the interactions required here, by means of digital quantum simulation techniques, as shown in Box~I. Each of the implemented time evolutions consists of a sequence of over 200 quantum gates (see Table I in Extended Data). In particular, in order to realise the non-local interactions $H_{zz}$ and $H_{\pm}$ with their
specific long-range interactions, we use global MS entangling gates together with a spectroscopic decoupling
method to tailor the range of the interaction. There, the population of the ions that are not involved
in the specific operations are shelved into additional electronic
states that are not affected by the light for the entangling
operations (see Methods). The local terms in $H_z$ correspond to Z rotations that are directly available in our set of
operations. The strength of all terms can be tuned by changing the
duration of the laser pulses corresponding to the physical operations.

Our scheme allows us to study a wide range of fundamental properties in 1D Wilson gauge theories. To demonstrate our approach, we concentrate on simulating the coherent quantum real-time dynamics of the Schwinger mechanism, i.e.\ the creation of particle--antiparticle pairs out of the bare vacuum $|\text{vacuum}\rangle$, the state where matter is entirely absent (see Methods). 
After initialising the system in this state, which corresponds to the ground state for $m\to\infty$ (see Fig.~\ref{Fig_2}(a)), we apply $\hat{H}_{\text{S}}$ (see Box~I(d)) for different masses and coupling strengths. As a first step, we measure the particle number density  $\nu(t)=\frac{1}{2N}\sum_{l=1}^N\langle(-1)^l \hat{\sigma}_l^z(t)+1\rangle$ generated after a simulated time evolution of duration $t$. The value $\nu=0.5$ corresponds to a state containing on average one pair (see Box~I(b)). 
As Fig.~\ref{Fig_2}(c) shows, an initial phase of rapid pair creation is followed by a reduction of $\nu(t)$ due to recombination effects. The measured evolution shows excellent agreement with theoretical predictions using a simple noise model for decoherence, as explained in the Methods. In Fig.~\ref{Fig_2}(b), we probe the particle-antiparticle generation for a broad range of masses $m$. Larger values of $m$ increase the energy cost of pair production and thus lead to faster oscillations with a suppressed magnitude (see also Methods and extended data).\\ 

Our platform allows for direct measurements of the vacuum persistence amplitude and of the generated entanglement. The vacuum persistence amplitude
$G(t)=\langle \text{vacuum}|e^{-i \hat{H}_{S}t}|\text{vacuum}\rangle$
quantitifies the decay of the unstable vacuum (see Methods).  
The associated probability $|G(t)|^2$ shown in Fig.~\ref{Fig_3}(a,c) is also known as the Loschmidt echo, which is important in other contexts such as the theory of quantum chaos~\cite{Gorin2006dv} and dynamical critical phenomena far from equilibrium~\cite{Heyl2013a}.

The vacuum decay continuously produces entanglement, as pairs are constantly generated and particles and antiparticles propagate away from each other, thus correlating distant parts of the system. Entanglement plays a crucial role in the characterisation of dynamical processes in quantum many-body systems, and its analysis permits us to quantify the quantum character of the generated correlations. 
To this end, we reconstruct the density matrix after each time step by full state tomography, and evaluate the entanglement of one half of the system with the other by calculating the logarithmic negativity. This quantity is an entanglement measure for mixed states~\cite{Plenio2007}, which is defined as the sum of the negative eigenvalues of the partially transposed density matrix. The entanglement between two contiguous blocks of our spin system is equivalent to the entanglement in the simulated fermionic system described by Eq.~\eqref{Eq_LatticeHamiltonian}, i.e.\ including the gauge fields (C.M \emph{et al.}, in preparation). 
In Fig.~\ref{Fig_3}(b,d), we show the real-time dynamics of the logarithmic negativity for different parameter regimes. 
Entanglement between the two halves of the system is due to the presence of a pair distributed across them. 
Accordingly, less entanglement is produced for increasing particle masses $m$ and field  energies $J$. 
The latter has a stronger influence, as it not only raises the energy cost for the creation of a pair but also for increasing the distance between particle and antiparticle.

Our study should be understood as a first step in the effort to simulate increasingly complex dynamics, including quantum simulations of lattice gauge theories~\cite{Preskill2012}, that cannot be tackled by classical numerical methods. 
Building on these results, future challenges include the quantum simulation of non-abelian lattice gauge theories and systems beyond 1D.

\appendix

\section{Methods}

\subsection{Encoding of the lattice Schwinger model in a spin model with long-range interactions}

Our starting point is the Kogut--Susskind Hamiltonian formulation of the lattice Schwinger model~\cite{Schwinger1,KogutSusskindFormulation}, see Eq.~(\ref{Eq_LatticeHamiltonian}) in the main text. This model describes one-component fermion fields $\hat{\Phi}_n$ that are located at lattice sites $n$ and interact with gauge fields that are represented by the canonically commuting operators $[\hat{\theta}_n,\hat{L}_m]=i\delta_{n,m}$ as illustrated in Box I. $\hat{\theta}_n$ and $\hat{L}_n$ represent the vector potential and electromagnetic field on the link connecting sites $n$ and $n+1$. The dynamics is constrained by the Gauss law
\begin{eqnarray}\label{Eq_GaussLaw_Lattice}
\hat{L}_{n}-\hat{L}_{n-1}=\hat{\Phi}^{\dag}_n\hat{\Phi}_n-\frac{1}{2}\left[1-(-1)^n\right].
\end{eqnarray}
Eq.~(\ref{Eq_GaussLaw_Lattice}) can be understood by considering a fixed field operator $\hat{L}_n$ and an adjacent spin $\hat{\Phi}_n$ to its right. As shown in Box~I(a), spins in state $|\!\!\uparrow\rangle$ ($|\!\!\downarrow\rangle$) on an odd (even) lattice site indicate that this lattice site is in the vacuum state, i.e.\ not occupied by a particle or antiparticle. Accordingly, $\hat{L}_{n}=\hat{L}_{n-1}$. Spins in the state $|\!\uparrow\rangle$ on even lattice sites (corresponding to positrons) generate $(+1)$ unit of electric flux to the right $\hat{L}_{n}=\hat{L}_{n-1}+1$. Similarly, spins in the state $|\!\downarrow\rangle$ on odd lattice sites (corresponding to electrons) lead to a decrease of one unit,  $\hat{L}_{n}=\hat{L}_{n-1}-1$. In order to cast the lattice Schwinger Hamiltonian given in  Eq.~(\ref{Eq_LatticeHamiltonian}) in the main text in the form of a spin model, the one-component fermion operators $\hat{\Phi}_n$ are mapped to Pauli spin operators by means of a Jordan-Wigner transformation~\cite{M_Banks1976},
\begin{eqnarray*}
\hat{\Phi}_n=\prod_{l<n}\left[i\hat{\sigma}_l^z\right]\hat{\sigma}^{-}_n, \ \ \ \  \hat{\Phi}^{\dag}_n=\prod_{l<n}\left[-i\hat{\sigma}_l^z\right]\hat{\sigma}^{+}_n.
\end{eqnarray*}
This leads to
\begin{eqnarray*}
\hat{H}_{\text{spin}}&=&w\sum_{n=1}^{N-1}\left[\hat{\sigma}_n^+
e^{i\hat{\theta}_n}\hat{\sigma}_{n+1}^{-}+\text{H.c.}\right]\\
&+&\frac{m}{2}\sum_{n=1}^N (-1)^n\hat{\sigma}_n^z+J\sum_{n=1}^{N-1}\hat{L}_n^2,
\end{eqnarray*}
where constant terms (energy offsets) have been omitted. Using this expression, the gauge degrees of freedom are eliminated in a two-step procedure~\cite{Encoding}. Firstly, the operators $\hat{\theta}_n$ are eliminated by a gauge transformation,
\begin{eqnarray*}
\hat{\sigma}^-_n \rightarrow \prod_{l<n}\left[e^{-i\hat{\theta}_l}\right]\hat{\sigma}^-_n.
\end{eqnarray*}
In a second step, the electric field operators $\hat{L}_n$ are eliminated iteratively using the spin version of the Gauss law given in Eq.~(\ref{Eq_GaussLaw_Lattice}),
\begin{eqnarray*}
\hat{L}_n-\hat{L}_{n-1}=\frac{1}{2}\left[\hat{\sigma}_n^z+(-1)^n\right].
\end{eqnarray*}
This yields the pure spin Hamiltonian which is realised in our simulation scheme,
\begin{eqnarray}\label{Eq_SchwingerHamiltonian}
\hat{H}_{S}&=&\frac{m}{2}\sum_{n=1}^{N}(-1)^n\hat{\sigma}_n^z+ w\sum_{n=1}^{N-1}\left[\hat{\sigma}_n^{+}\hat{\sigma}_{n+1}^{-}+\text{H.c.}\right]\nonumber\\
&+&J\sum_{n=1}^{N-1}\left[\epsilon_0+\frac{1}{2}\sum_{m=1}^n\left[\hat{\sigma}_m^z+(-1)^m\right]\right]^2.
\end{eqnarray}
The free parameter $\epsilon_0$ corresponds to the boundary electric field on the link to the left of the first lattice site (see Box~I(b,c)). Throughout this paper we consider the case of zero background field, where $\epsilon_0=0$.\\ 

The gauge fields do not appear explicitly in this description. Instead, they effectively generate a non-local long-range interaction that corresponds to the Coulomb interaction between the simulated charged particles. 
So far, the encoding approach to the Schwinger model that has been explained in this section, has been only been employed as a tool for analytical or numerical calculations~\cite{Encoding,M_MC1,M_MC2}. In contrast, we investigate here the use of this idea for a quantum simulation scheme, i.e.\ the realisation of the Schwinger model in its encoded form in an actual physical system. 
\subsection{Digital quantum simulation of the encoded Schwinger model}
We realise $\hat{H}_S$ given in Eq.~(\ref{Eq_SchwingerHamiltonian}) by means of a digital quantum simulation scheme~\cite{Lloyd96}, which will be described in detail in a manuscript in preparation by C.M. \emph{et al.}. For convenience, we express the simulated Hamiltonian in the form
\begin{eqnarray}\label{EqM_Hsim}
\hat{H}_S&=&\hat{H}_{ZZ}+\hat{H}_{\pm}+\hat{H}_Z,
\end{eqnarray}
where the three  parts of the Hamiltonian correspond to the two different types of two-body couplings $\hat{H}_{ZZ}$ and $\hat{H}_{\pm}$, as well as local terms $\hat{H}_Z$,
\begin{eqnarray*}
\hat{H}_{ZZ}&=&J\sum_{n< m}c_{nm}\hat{\sigma}^z_{n}\hat{\sigma}^z_m,\\
\hat{H}_{\pm}&=&w\sum_{n}\left(\hat{\sigma}_n^{+}\hat{\sigma}_{n+1}^{-}+\hat{\sigma}_{n+1}^{+}\hat{\sigma}_n^{-}\right),\\
\hat{H}_{Z}&=&m\sum_{n}c_n\hat{\sigma}_n^{z}+J\sum_{n}\tilde{c}_n\hat{\sigma}_n^{z}.
\end{eqnarray*}
The simulation protocol is based on time-coarse graining, where the desired dynamics of the Hamiltonian given by Eq.~(\ref{Eq_SchwingerHamiltonian}) is obtained whithin a time-averaged description. As illustrated in Box~I(f), the total simulation time $t_{\text{sim}}$ is divided into individual time windows of duration $T$. During each of these time windows, a full cycle of the protocol that is described below is performed. This cycle is repeated multiple times from $t=0$ to $t=t_{\text{sim}}$ and consists of three sections, as shown in Box~I(g). Each of these sections corresponds to one of the three parts of the desired Hamiltonian given by Eq.~(\ref{EqM_Hsim}). In the first section, $\hat{H}_{ZZ}$ is simulated, in the second, the nearest neighbour terms $\hat{H}_{\pm}$ are realised and in the third, the single particle rotations $\hat{H}_Z$ are performed.
In this way, the simulation scheme uses only two types of interactions, local rotations and an infinite-range entangling operation 
\begin{eqnarray}\label{Eq_HMS_X}
\hat{H}_{\text{MS\tiny{X}}}=J_0\sum_{n,m}\hat{\sigma}_n^x \hat{\sigma}_m^x,
\end{eqnarray}
which is routinely implemented in trapped ions by means of M\o lmer-S\o rensen gates~\cite{MSgates}.
In the following, we explain how the individual parts of the Hamiltonian are realised. More detailed explanations can be found in a manuscript in preparation by C.M. \emph{et al.}. The  relative strengths of the individual parts of $\hat{H}_S$, $J$, $w$ and $m$, can be tuned by adjusting the length of the elementary time windows or the strength of the underlying interaction $J_0$ accordingly.

\subsubsection{Long-range interactions $\hat{H}_{ZZ}$}

The first part of Eq.~(\ref{EqM_Hsim}) originates from the third term in Eq.~(\ref{Eq_SchwingerHamiltonian}) representing the electric-field energy. It takes the form
\begin{eqnarray}\label{Eq_Hzz}
\hat{H}_{ZZ}=\frac{J}{2}\sum_{m=1}^{N-2}\sum_{n=m+1}^{N-1}(N-n)\hat{\sigma}_m^{z} \hat{\sigma}_n^z,
\end{eqnarray}
and describes two-body interactions with an asymmetric distance dependence, where each spin interacts with constant strength with all spins to its left, while the coupling to the spins on its right decreases linearly with distance (see Box~I(d,e)). As the number of elements in the  spin coupling matrix is proportional to $N^2$, a brute force digital simulation approach to this problem would require $N^2$ time steps. Using our protocol, which is inspired by techniques put forward in~\cite{M_StrobShake}, the required resources scale only linearly in $N$. This is accomplished using the scheme illustrated in Box~I(h). We introduce $N-2$ time windows, which can be shown to be the minimal number of time steps required to simulate the Hamiltonian in Eq.~(\ref{Eq_Hzz}). Each elementary time window has length $\Delta t_I$. In the $n^{\text{th}}$ time window, the Hamiltonian 
\begin{eqnarray*}
\hat{H}_{\text{MS\tiny{Z}}}^{(n)}=J_0\sum_{i,j}^{n+1}\hat{\sigma}_i^z\hat{\sigma}_j^z
\end{eqnarray*}
is applied. $\hat{H}_{\text{MS\tiny{Z}}}^{(n)}$ is realised by applying the Hamiltonian given in Eq.~(\ref{Eq_HMS_X}) in combination with local rotations, $R(y)\hat{H}_{\text{MS\tiny{X}}}R^\dag(y)=\hat{H}_{\text{MS\tiny{Z}}}$, where $R(y)=e^{i\frac{\pi}{4}\sum_{i=1}^N \hat{\sigma}_i^y}$. The resulting time-averaged Hamiltonian for the first section of the time interval $T$, $\bar{H}_{I}=\frac{1}{N-2}\sum_{n=1}^{N-2}\hat{H}_{\text{MS\tiny{Z}}}^{(n)}$ is proportional to the desired Hamiltonian in Eq.~(\ref{Eq_Hzz}), $\bar{H}_{I}=\frac{2}{N-2}\frac{J_0}{J}H_{ZZ}$.

As shown in Box~I(h), only ions $1$ to $n+1$ participate in the entangling interaction in time step $n$. Since the interaction is implemented via a global beam that couples to the entire ion string {(see Fig.~\ref{Fig_1}(b))}, ions $n+2$ to $N$ are decoupled by applying hiding pulses. The population in the qubit {states} of these ions is transferred to electronic levels that are not affected by the interaction using suitable laser pulses. The population in the state $\ket{\downarrow} = 4S_{1/2} \ (\text{magnetic number } m = -1/2)$ is transferred to the state $3D_{5/2} \ (m = -5/2)$, and the population in $\ket{\uparrow} = 3D_{5/2} \ (m = -1/2)$ is transferred to the state $3D_{5/2} \ (m = -3/2)$ via $4S_{1/2} \ (m = +1/2)$.

\subsubsection{Nearest neighbour terms $\hat{H}_{\pm}$}

The second part of Eq.~(\ref{EqM_Hsim}),
\begin{eqnarray*}
\hat{H}_{\pm}=w\sum_{n=1}^{N-1}\left(\hat{\sigma}_n^{+}\hat{\sigma}_{n+1}^{-}+\text{H.c.}\right)
\end{eqnarray*}
corresponds to the creation and annihilation of particle-antiparticle pairs (see Box~I(a,c)). For realising this Hamiltonian, the interaction given in Eq.~(\ref{Eq_HMS_X}) needs to be modified not only in range, but also regarding the type of coupling. This is accomplished by dividing the time window dedicated to realising $\hat{H}_{\pm}$ (see Box~I(g)) into $N-1$ elementary time slots of length $\Delta t_{II}$. Each of these is used for inducing the required type of interaction between a specific pair of neighbouring ions. For example, the first elementary time slot of length $\Delta t_{II}$ is used to engineer an interaction of the type $\hat{H}_{ij}\propto \hat{\sigma}{_i}^+\sigma_j^-+{\text{H.c.}}$ between the first and the second spin, the second time slot is used to do the same for the second and the third spin, and so on. This can be done by applying suitable hiding pulses, to all spins except for a selected pair of  ions $i$ and $j$. The selected pair undergoes a sequence of gates, which transforms the $\hat{\sigma}^x_i\hat{\sigma}_j^x$-type coupling in Eq.~(\ref{Eq_HMS_X}) into an interaction of the required form  and consists of four steps: 
(i) a single qubit operation on the two selected spins $i$ and $j$, $U=e^{i \frac{\pi}{4}(\hat{\sigma}^z_i+\hat{\sigma}_j^z)}$ 
(ii) an evolution under the Hamiltonian given in Eq.~(\ref{Eq_HMS_X}) for the selected pair of spins, $\hat{H}_{\text{MS\tiny{X}}}^{(ij)}$ during a time $\Delta t _{II}/2$, $e^{-i\hat{H}_{\text{MS\tiny{X}}}^{(ij)}\Delta t_{II}/2}$ 
(iii) another single qubit operation $U^{\dag}$ and finally 
(iv) another two-qubit gate $e^{i\hat{H}_{\text{MS\tiny{X}}}^{(ij)}\Delta t_{II}/2}$. 
The time evolution operator associated with the described sequence of gates is given by $e^{i H_{II} ^{(ij)}\Delta t_{II}}$ with 
\begin{eqnarray*}
\hat{H}_{II}^{(ij)}&=&\frac{1}{2}\left(\hat{H}_{\text{MS\tiny{X}}}^{(ij)}+ U^{\dag} \hat{H}_{\text{MS\tiny{X}}}^{(ij)}U\right)=J_0\left(\hat{\sigma}_i^{+}\hat{\sigma}_j^- +\text{H.c.}\right),
\end{eqnarray*}
as desired. The relative strength of the nearest neighbour terms $\hat{H}_{\pm}$ and the long-range
couplings $\hat{H}_{ZZ}$, $w/J$ can be adjusted by tuning the ratio of the lengths of the elementary time windows $\Delta_{II}/\Delta_{I}$.

\subsubsection{Single-particle terms $\hat{H}_Z$}

The last contribution to the Hamiltonian in Eq.~(\ref{EqM_Hsim}) consists of two terms $\hat{H}_{Z}=m\sum_{n}c_n\hat{\sigma}_n^{z}+J\sum_{n}\tilde{c}_n\hat{\sigma}_n^{z}$. The first term in this expression reflects the rest masses of the fermions. The second term is an effective single-particle contribution originating from the third part of Eq.~(\ref{Eq_SchwingerHamiltonian}) and corresponds to a change in the effective fermion masses due to the elimination of the electric fields. The local terms of the simulated Hamiltonian are given by
\begin{eqnarray*}
\hat{H}_{Z} = \frac{m}{2}\sum_{n=1} ^N(-1)^n\hat{\sigma}_n^z-\frac{J}{2}\sum_{n=1}^{N-1} (n \operatorname{mod}2)\sum_{l=1}^n\hat{\sigma}_l^z.
\end{eqnarray*}
These are implemented by means of AC--Stark shifts, induced by laser pulses that are far red-detuned from the qubit transition~\cite{Blatt2012,Schindler2013}.

\subsection{Measurement and postselection}

For each set of system parameters and number of simulation time steps, we perform a full state tomography to determine the density matrix that corresponds to the quantum state of the system. The electronic state of the ions is detected via a fluorescence measurement using the electron shelving technique \cite{Schindler2013}. The entire string is imaged {by} a CCD camera, performing a full projective measurement in the Z basis. This procedure is repeated 100 times to gather sufficient statistics.

As a consequence of charge conservation, an equal number of particles and antiparticles {is} created during the ideal dynamics of the system. Since our evolution starts with the vacuum state, the physical Hilbert space of the simulation is spanned by the six states $\{ \ket{0000} = \ket{\uparrow \downarrow \uparrow \downarrow}$, $\ket{\text{e}^-\text{e}^+00} = \ket{\downarrow \uparrow \uparrow \downarrow}$, $\ket{0\text{e}^+\text{e}^-0} = \ket{\uparrow \uparrow \downarrow \downarrow}$, $\ket{00\text{e}^-\text{e}^+} = \ket{\uparrow \downarrow \downarrow \uparrow}$, $\ket{\text{e}^-00\text{e}^+} = \ket{\downarrow \downarrow \uparrow \uparrow}$, {and} $ \ket{\text{e}^-\text{e}^+\text{e}^-\text{e}^+} = \ket{\downarrow \uparrow \downarrow \uparrow} \}$, where $\ket{0}$ denotes the vacuum, $\ket{e^-}$ a particle and $\ket{e^+}$ an antiparticle. However, experimental errors during the simulation produce leakage from this subspace, such that nonphysical states such as  $\ket{\text{e}^-000} = \ket{\downarrow \downarrow \uparrow \downarrow}$ get populated. Therefore, the raw measured density matrices $\rho_{\text{raw}}$ are projected onto the Hilbert space spanned by the physical states and normalized,
\begin{linenomath*}
\begin{equation*}
\rho_{\text{phys}} = \frac{P \rho_{\text{raw}} P}{\operatorname{tr}(P \rho_{\text{raw}} P)},
\end{equation*}
\end{linenomath*}
where $P$ is the projector onto the physical subspace. All experimental data presented in this work correspond to physical density matrices $\rho_{\text{phys}}$ postselected in this way.

A useful measure of the fidelity of the evolution is the population leakage from the physical subspace. After $\{1,2,3,4\}$ evolution time steps, the measured populations remaining in the physical subspace were on average $\{86 \pm 2, 79 \pm 1, 73 \pm 1, 69 \pm 1\}$\% {of the population} before postselection.

\subsection{Experimental error model}

To gauge the performance of the experimental simulation and to understand the origin of the remaining sources of errors, we compare the experimental data to a simple phenomenological error model. Since the postselection already partially corrects for population errors, we considered an error model that 
{describes}
uncorrelated dephasing, parameterised with an phase flip error probability $p$ per qubit and per evolution time step. The density matrix $\rho$ is then, at each evolution step, subject to the composition of the error channels $\mathcal{E}_i$ for each qubit,

\begin{align*}
\rho &\rightarrow \mathcal{E}_4 \circ \mathcal{E}_3 \circ \mathcal{E}_2 \circ\mathcal{E}_1(\rho), \ \text{where:}\\
\mathcal{E}_i(\rho) &= (1 - p) \rho + p \sigma^z_{i} \rho \sigma^z_{i}.
\end{align*}
The value for the error probability $p$ was extracted from a fit to all of the experimental data collected. For all the data taken with nonzero $J$ we found a value of $p = 0.038$. Whenever $J = 0$, the simulation does not require any ZZ interactions. Thus, several entangling gates are {omitted} from the sequence and consequently higher fidelities are expected. Indeed, for this case the error probability per time step was found to be $p = 0.031$.

\subsection{Quantum simulation of the Schwinger mechanism}
We simulate the coherent real-time dynamics in the Schwinger model focusing on the Schwinger mechanism, i.e. spontaneous particle-antiparticle production out of the unstable vacuum.
This effect is at the heart of quantum electrodynamics and its observation is currently pursued at high intensity laser facilities ELI and XCELS~\cite{Narozhny2014} (theoretical proposals for its quantum simulation can for example be found in~\cite{Wiese:2013kk,Zohar2015,Alvarez2015,Casanova2011}).
To simulate the dynamics of pair creation, we consider as is usual~\cite{calzetta_book,Hebenstreit2013} the bare vacuum as initial state, where matter is completely absent, $\ket{\mathrm{vacuum}} = \ket{0000}$. In the spin representation this state is given by $ |\uparrow \downarrow \uparrow \downarrow\rangle$ accordingly. Note that the bare vacuum is different from the so-called dressed vacuum state, which is the ground state of the full Hamiltonian.

\subsubsection{Decay of the vacuum}

The natural quantity characterising the decay of the unstable vacuum is the vacuum persistence amplitude introduced by J.\ Schwinger~\cite{M_Schwinger1951}, which is defined as the overlap of the initial state $|\Psi(0)\rangle=|\mathrm{vacuum}\rangle$ with the time-evolved state
\begin{linenomath*}
\begin{equation*}
	G(t) = \langle \mathrm{vacuum} | e^{-i\hat{H}_St} | \mathrm{vacuum} \rangle.
\end{equation*}
\end{linenomath*}
Within the original formulation, the Schwinger mechanism was considered for
the continuum system and a classical electric field of strength $E$~\cite{M_Schwinger1951}. There,
it has been shown that the particle number density $\nu(t)$ is directly related
to the rate function $\lambda(t)$, that characterises the decay of the vacuum persistence probability $\big| G(t)\big|^2$,
\begin{linenomath*}
\begin{equation*}
\lambda(t) = - \lim_{N\to\infty} \frac{1}{N} \log\Big[ \big| G(t)
\big|^2 \Big].
\end{equation*}
\end{linenomath*}
Specifically, in the limit of large fermion masses $m \gg \sqrt{qE}$ with $q$ the electric
charge, as relevant in the high-energy context, $\lambda(t) =
\nu(t)$ for thermodynamically large systems in the continuum.

Since vacuum persistence amplitudes have so far not been measured, this connection between $\lambda(t)$ and $\nu(t)$ has not yet been tested experimentally. In Extended Data Figure 1, we show the measured rate function $\lambda(t)$ and find good qualitative agreement with $\nu(t)$, even for the few qubits in our digital quantum simulation. 

\subsubsection{Finite size effects}
In the following, we discuss the dependence of the results on the number of lattice sites $N$.  Extended Data Figure 2 shows the time evolution of the particle number density and the entanglement for different system sizes $N$. 
Already for our experimental system with $N=4$, we find qualitative agreement with respect to the results expected for larger $N$. By scaling up the system, the dynamics quickly converges for the considered parameters. We address the continuum limit $a\rightarrow 0$, $N\rightarrow \infty$ for fixed values of the coupling $g$ and the mass $m$ in a manuscript in preparation (C.M. \emph{et al.}).

\section{Acknowledgements}

The authors would like to acknowledge discussions with Cornelius Hempel and Enrique Rico Ortega. Financial support was provided by the Austrian Science Fund (FWF), through the SFB FoQuS (FWF Project No. F4002-N16 and No. F4016-N23), by the European Commision via the integrated project SIQS and the ERC synergy grant UQUAM, by the Deutsche Akademie der Naturforscher Leopoldina (Grant No. LPDS 2013-07 and No. LPDR 2015-01), as well as the Institut f\"ur Quantenoptik und Quanteninformation GmbH. E.A.M. is a recipient of a DOC fellowship from the Austrian Academy of Sciences. P.S. was supported by the Austrian Science Foundation (FWF) Erwin Schr\"odinger Stipendium 3600-N27. This research was funded by the Office of the Director of National Intelligence (ODNI), Intelligence Advanced Reasearch Projects Activity (IARPA), through the Army Research Office grant W911NF-10-1-0284. All statements of fact, opinion or conclusions contained herein are those of the authors and should not be construed as representing the official views or policies of IARPA, the ODNI, or the U.S. Government.

\section{Author contributions}

E.M., C.M., M.D. and T.M. developed the research based on discussions with P.Z. and R.B.; E.M. and P.S. performed the experiments. E.M., C.M., P.S. and M.H. analysed the data and carried out numerical simulations. E.M., P.S., D.N., A.E. and T.M. contributed to the experimental setup. C.M., M.H., M.D., P.H. and P.Z. developed the theory. E.M., C.M., P.S., M.H., P.H., M.D., P.Z. and R.B. wrote the manuscript and provided revisions. All authors contributed to discussions of the results and the manuscript.

%\section{Author information}

%Reprints and permissions information is available at www.nature.com/reprints. The authors declare no competing financial interests. Correspondence and requests for materials should be addressed to Esteban A. Martinez (esteban.martinez@uibk.ac.at).

\clearpage

\section{Extended data}

\setcounter{figure}{0}

\begin{figure}[h]
\centering
\includegraphics[width=\columnwidth]{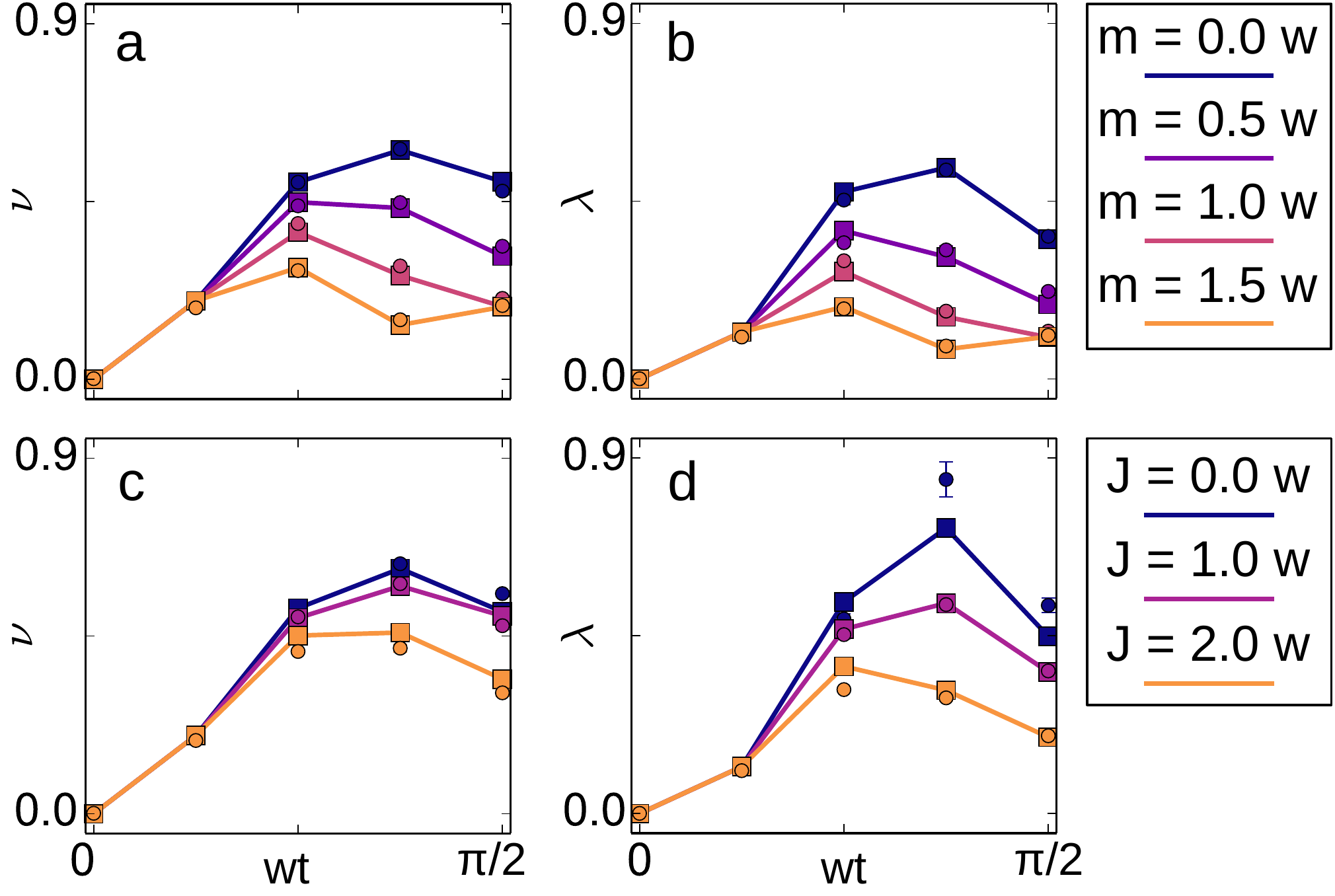}
\caption{\label{fig:lambdas} Evolution of the particle number density $\nu(t)$ and the rate function $\lambda(t)$, which characterises the decay of the vacuum persistence probability 
$\big| G(t)\big|^2 = e^{-N\lambda(t)}$. Panels (a) and (b) show the time evolution of these quantities for different values of the particle mass $m$ and fixed electric field energy $J=w$, where $w$ is the rate of particle-antiparticle creation and annihilation (see Eq.~(\ref{Eq_LatticeHamiltonian}) in the main text). Panels (c) and (d) show the evolution of $\nu(t)$ and $\lambda(t)$ for different values of $J$ and fixed particle mass $m=0$ as a function of the dimensionless time $w t$.}
\end{figure}

\begin{figure}[h]
\includegraphics[width=\columnwidth]{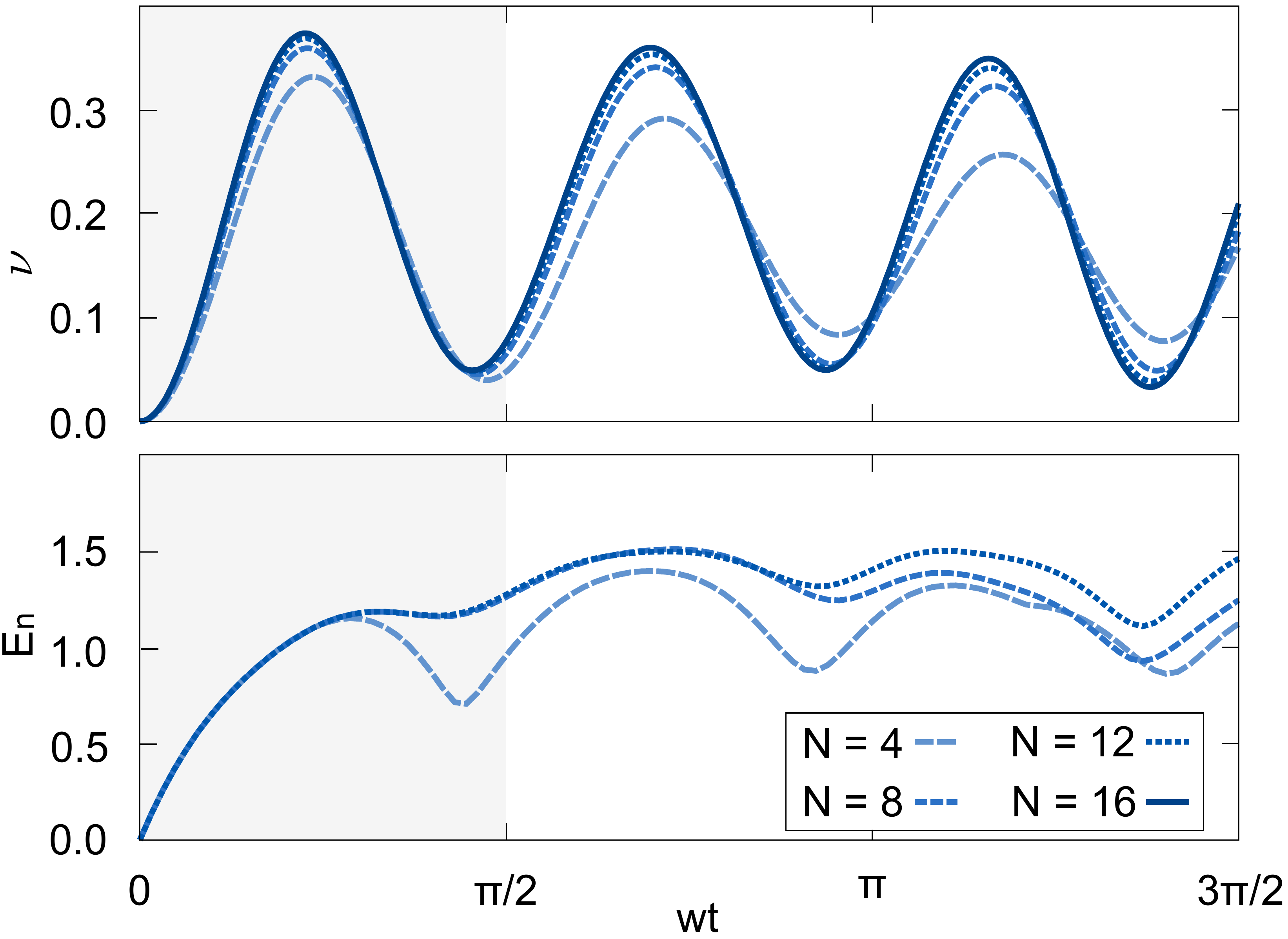}
\caption{Evolution of the particle number density $\nu=\frac{1}{2N}\sum_{l=1}^N\langle(-1)^l \hat{\sigma}_l^z(t)+1\rangle$  and the logarithmic negativity $E_n$ for for different system sizes $N$. The logarithmic negativity is evaluated with respect to a cut in the middle of the considered spin chain und quantifies the entanglement between the two halves of the system. Both quantities are shown as a function of the dimensionless time $wt$ for $J=m=w$.}
\end{figure}

\begin{table*}[h]
\includegraphics[width=\linewidth]{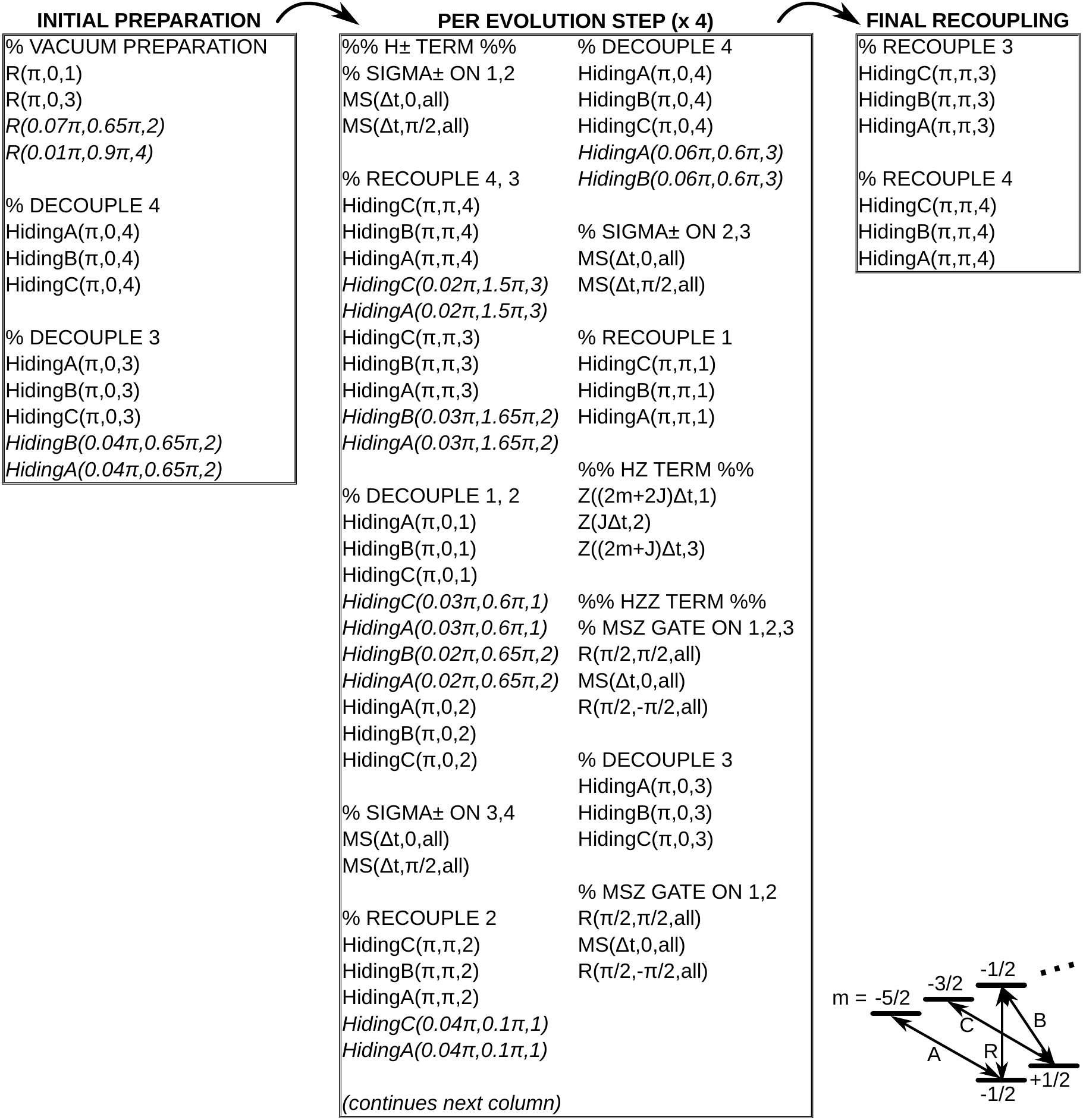}
\caption{Pulse sequence performed in the experiment. The pulses are listed in the order in which they are applied. The operations shown in the middle box are repeated once per evolution step, resulting in a total number of $12 + 51 \times 4 + 6 = 222$ pulses for 4 evolution steps. The pulses are labeled in the form $\operatorname{Pulse}(\theta, \phi, \text{target qubit})$, where $\theta$ is the rotation angle (length) of the pulse, $\phi$ its phase, and the target qubit is an integer from 1 to 4 for addressed operations or ``all'' for global operations. ``R'' denotes a pulse on the qubit transition $4S_{1/2} (m = -1/2)$ to $3D_{5/2} (m = -1/2)$. ``MS'' corresponds to a M{\o}lmer-S{\o}rensen gate on the same transition. The hiding pulses ``HidingA,B,C'' are applied on the transitions: A) $4S_{1/2} (m = -1/2)$ to $3D_{5/2} (m = -5/2)$, B) $4S_{1/2} (m = +1/2)$ to $3D_{5/2} (m = -1/2)$, C) $4S_{1/2} (m = +1/2)$ to $3D_{5/2} (m = -3/2)$. The pulses shown in italics serve the purpose of correcting addressing crosstalk.}
\end{table*}

\end{document}